\makeatletter\@ifundefined{selectfont}{%
\documentstyle[12pt]{article}}
{\documentstyle[12pt,oldlfont]{article}}
\makeatother
\begin{document}\thispagestyle{empty}\begin{flushright}
OUT--4102--49\\
INP--94--1/323\\
BI--TP/94--02\\
MPI--PhT/94--2\\
11 March, 1994                                      \end{flushright}
                                                    \vspace*{2mm}
                                                    \begin{center}
                                                    {\large\bf
Two-loop gluon-condensate contributions to heavy-quark\\[4pt]
current correlators: exact results and approximations$^{*)}$}
                                                    \vglue 5mm{\bf
D.~J.~Broadhurst$^{1)}$                             }
                                                    \vglue 2mm
Physics Department, Open University,
Milton Keynes, MK7 6AA, UK
                                                    \vglue 4mm{\bf
P.~A.~Baikov$^{2)}$, V.~A.~Ilyin$^{3)}$             }
                                                    \vglue 2mm
Nuclear Physics Institute of Moscow State University,
119 899 Moscow, Russia
                                                    \vglue 4mm{\bf
J.~Fleischer$^{4)}$, O.~V.~Tarasov$^{5)}$           }
                                                    \vglue 2mm
Fakult\"{a}t f\"{u}r Physik, Universit\"{a}t
Bielefeld, D-33615 Bielefeld 1, Germany
                                                    \vglue 4mm{\bf
V.~A.~Smirnov$^{6)}$                                }
                                                    \vglue 2mm
Max-Planck-Institut f\"{u}r Physik,
Werner-Heisenberg-Institut, D-80805 Munich, Germany \end{center}
                                                    \vfill
{\bf Abstract}
The coefficient functions of the gluon condensate $\langle G^2\rangle$, in
the correlators of heavy-quark vector, axial, scalar and pseudoscalar
currents, are obtained analytically, to two loops, for all values of
$z=q^2/4m^2$. In the limiting cases $z\to0$, $z\to1$, and $z\to-\infty$,
comparisons are made with previous partial results. Approximation methods,
based on these limiting cases, are critically assessed, with a view to
three-loop work. High accuracy is achieved using a few moments as input. A
{\em single\/} moment, combined with only the {\em leading\/} threshold and
asymptotic behaviours, gives the two-loop corrections to better than 1\% in
the next 10 moments. A two-loop fit to vector data yields
$\langle\frac{\alpha_{\rm s}}{\pi}G^2\rangle\approx0.021$~GeV$^4$.
                                                    \vfill
                                                    \footnoterule\noindent
$^*$) Collaboration supported in part by INTAS grant 93 1180\\
$^1$) D.Broadhurst@open.ac.uk\\
$^2$) Baikov@theory.npi.msu.su\\
$^3$) Ilyin@theory.npi.msu.su;
      supported in part by RFFR grant N~93-02-14428\\
$^4$) Fleischer@physik.uni-bielefeld.de\\
$^5$) oleg@physik.uni-bielefeld.de;
      supported by BMFT and RFFR grant N~93-02-14428;\\\phantom{$^*$)}
      on leave of absence from JINR, 141 980 Dubna, Russia\\
$^6$) vls@dmumpiwh.bitnet;
      supported by the Humboldt Foundation;\\\phantom{$^*$)}
      on leave of absence from NPI, Moscow State University,
      119 899 Moscow, Russia
\newpage
\setcounter{page}{1}
\newcommand{\df}[2]{\mbox{$\frac{#1}{#2}$}}
\newcommand{\Eq}[1]{~(\ref{#1})}
\newcommand{\Eqq}[2]{~(\ref{#1},\ref{#2})}
\newcommand{\vev}[1]{\langle#1\rangle}
\newcommand{\MS}{\mbox{$\overline{\rm MS}$}}
\newcommand{\as}{\alpha_{\rm s}}
\newcommand{\rd}{{\rm d}}
\newcommand{\re}{{\rm e}}
\newcommand{\ri}{{\rm i}}
\newcommand{\rT}{{\rm T}}
\newcommand{\rV}{^{\rm V}}
\newcommand{\rA}{^{\rm A}}
\newcommand{\rS}{^{\rm S}}
\newcommand{\rP}{^{\rm P}}
\newcommand{\np}{_{\rm np}}
\newcommand{\Li}{{\rm Li}_2}
\newcommand{\rSP}{^{\rm S,P}}
\newcommand{\VASP}{{\rm V,A,S,P}}
\newcommand{\Pm}{\phantom{-}}
\newcommand{\pfrac}[2]{\Pm\frac{#1}{#2}}
\newcommand{\qP}{{}\quad P}
\newcommand{\hl}{\\\hline&&&&&&&\\[-12pt]}
\newcommand{\noshow}{\mbox{input}}
\section{Introduction}
Whilst considerable progress on multi-loop diagrams was made between the
AI-HENP 92~\cite{AI92} and AI-HENP 93~\cite{AI93} workshops, there is still
a pressing need for methods that exploit hard-won analytical results by
efficient techniques of numerical approximation. Progress in this direction
was made, independently, in~\cite{YF} and~\cite{BFT}, whose authors combine
in this present work, which derives new analytical two-loop results for
heavy-quark current correlators and uses them to refine the
phenomenological extraction~\cite{YF} of the gluon condensate and to assess
the numerical methods of~\cite{YF,BFT}.
\par
In Section~2, we obtain analytically, to two loops, the coefficient
functions of the gluon condensate, $\langle G^2\rangle$, in the correlators
of heavy-quark vector, axial, scalar and pseudoscalar currents, for all
values of $z=q^2/4m^2$. This work is therefore an extension of that
in~\cite{PL1,PL2}, where the first 7 moments of these coefficient functions
were analytically computed, by reducing the problem to the evaluation of
vacuum scalar two-loop diagrams with one massless and two massive lines
and arbitrary integer indices. Here we apply the more general strategy
of~\cite{BFT,MAS}, reducing the problem to evaluation of two-point
diagrams, of the type needed for the two-loop photon propagator~\cite{B+R},
and using the programs developed for~\cite{BFT} to obtain explicit
analytical expressions for such diagrams, in $d=4-2\varepsilon$ spacetime
dimensions. As input to these programs, we use the intermediate results
of~\cite{PL1,PL2}, obtained after the calculation of traces and the action
of the projector for the operator $G^2$.
\par
Renormalization of the resulting bare two-loop terms amounts to no more
than mass renormalization of the $d$\/-dimensional one-loop contributions,
which we perform in the on-shell scheme, expressing our results in terms of
the pole-mass, $m$. Doing this, we discover that a term was missed
in~\cite{PL1,PL2}, when performing the limit $\varepsilon\to0$ in the \MS\
scheme, with the effect that the results given in~\cite{YF,PL1,PL2} do not
hold in the \MS\ scheme (nor in any scheme that respects the Ward identity
relating axial and pseudoscalar correlators). After correcting this
inconsistency, we find agreement between our new results and the partial
results of~\cite{PL1,PL2} for the first 7 moments. In addition to studying
this $z\to0$ limit in Section~3, we also compare our results with those
obtained in~\cite{LG2}, as $z\to-\infty$, and in~\cite{NR}, as $z\to1$,
finding agreement with the leading term, in each case.
\par
In Section~4, we assess numerical approximations, developed
in~\cite{AI93,YF,BFT,DST,CONF}, by comparing predictions, made on the basis
of previously available input~\cite{PL1,LG2,NR}, with exact new two-loop
results. The outcome is most satisfactory: with 7 moments as input, the
methods achieve very high accuracy; with 4 moments, they fare almost as
well; most remarkably, we shall show that a {\em single\/} moment, combined
with only the {\em leading\/} behaviours as $z\to1$ and $z\to-\infty$,
gives the two-loop terms in the next 10 moments to better than 1\%
accuracy.
\par
Section~5 gives our conclusions, concerning existing phenomenology and
future calculations, both analytical and numerical. One significant
motivation for studying numerical approaches with limited analytical input
concerns the feasibility of approximating the three-loop photon propagator
from a few of its moments, obtainable by the methods of~\cite{3LP}, and then
using it for applications such as the muon anomaly~\cite{GM2}. We suggest
that such an approach to ambitious calculations in both QED and QCD is
indeed feasible.
\newpage
\section{Exact results}
We evaluate contributions to the correlators, $\ri\int\re^{\ri q
x}\vev{\rT(J(x)J(0))}\rd x$, of the vector (V) current
$J\rV_\mu=\overline\psi\gamma_\mu\psi$, the axial (A) current
$J\rA_\mu=\overline\psi\gamma_\mu\gamma_5\psi$, the scalar (S) current
$J\rS=2m\overline\psi\psi$, and the pseudoscalar (P) current
$J\rP=2m\overline\psi\ri\gamma_5\psi =\partial^\mu J\rA_\mu$, where $\psi$
is a heavy-quark field of mass $m$. We choose to work in the on-shell (OS)
scheme, where the results are simplest. Since the currents have no
anomalous dimensions, one may translate the results to any other scheme
(e.g.\ the \MS\ scheme) by making a one-loop transformation from the pole
mass, $m$, to the renormalized mass of that scheme (e.g.\ by using
$m=\overline{m}(\mu)(1+(\ln(\mu^2/m^2)+\frac43)\as/\pi +O(\as^2))$, where
$\overline{m}(\mu)$ is the \MS\ mass, at scale $\mu$).
\par
We denote the correlators of $J\rSP$ by $\Pi\rSP(q^2)$ and decompose the
tensor structure of the vector and axial correlators
as follows:
\begin{eqnarray}
\ri\int\re^{\ri q x}\vev{\rT(J\rV_\mu(x)J\rV_\nu(0))}
\rd x&=&\Pi\rV(q^2)(q_\mu q_\nu-q^2g_{\mu\nu})\,,
\label{piv}\\
\ri\int\re^{\ri q x}\vev{\rT(J\rA_\mu(x)J\rA_\nu(0))}
\rd x&=&\Pi\rA(q^2)\left(\frac{q_\mu q_\nu}{q^2}-g_{\mu\nu}\right)
+\frac{\Pi\rP(q^2)-\Pi\rP(0)}{q^4}q_\mu q_\nu\,,
\label{pia}
\end{eqnarray}
where $\Pi\rP(0)=-4\vev{m\overline\psi\psi}$ enters\Eq{pia} as an
equal-time commutator~\cite{DJB,SCG}. Finally, we define dimensionless
coefficients of the non-perturbative gluon condensate
by writing the dimension-4 contributions to the correlators as
\begin{equation}
\Pi^J\np(q^2)=\frac{\vev{(\as/\pi)G_{\mu\nu}^a G^{\mu\nu}_a}}
{(2m)^{n_J}}\left(C^J(z)+O(\as^2)\right)\,;\quad
C^J(z)=C^J_1(z)+\frac{\as}{\pi}C^J_2(z)\,,
\label{cj}
\end{equation}
in the channels $J=\VASP$, with exponents $n_J=4,2,0,0$, respectively,
making $C^J(z)$ a dimensionless function of $z\equiv q^2/4m^2$, with one-
and two-loop contributions $C^J_{1,2}(z)$. Since the currents are not
renormalized, and $\as G^2$ is not renormalized at one loop, the
renormalization scale of $\as$ in the two-loop term is irrelevant here. It
should, however, be taken as $O(m)$, to suppress large logarithms at
three-loop order~\cite{3LP}.
\par
Using {\sc Reduce}~\cite{RED} for the trace calculations of~\cite{PL1,PL2}
and for the recursion of the resulting scalar integrals of~\cite{BFT}, we
obtained the following on-shell results in the 4 channels:
\begin{eqnarray}
48z^2(1-z)^2C\rV(z)&=&-3+4z-4z^2+3(1-2z)G(z)+\frac{\as}{\pi}P\rV(z)\,,
\label{vres}\\
8z(1-z)C\rA(z)&=&\Pm1-2z-G(z)+\frac{\as}{\pi}P\rA(z)\,,
\label{ares}\\
16z(1-z)C\rS(z)&=&-1-2z+(1-4z)G(z)+\frac{\as}{\pi}P\rS(z)\,,
\label{sres}\\
48(1-z)^2C\rP(z)&=&\Pm7-10z+3(3-4z)G(z)+\frac{\as}{\pi}P\rP(z)\,,
\label{pres}\\
P^J(z)&=&P^J_1(z)+P^J_2(z)G(z)+P^J_3(z)G^2(z)+(1-z)P^J_4(z)H(z)\,,
\label{pj}
\end{eqnarray}
with polynomials, $P^J_i(z)$, given in Table~1. As $d\to4$, two basic
integrals are encountered:
\begin{eqnarray}
G(z)&=&\frac{2y}{y^2-1}\ln y\,;\quad
y=\frac{\sqrt{1-1/z}-1}{\sqrt{1-1/z}+1}\,,
\label{gz}\\
H(z)&=&\frac{2y}{y^2-1}\left(\Li(y^2)-\Li(y)
+(2\ln(1+y)+\ln(1-y)-\df34\ln y)\ln y\right),
\label{hz}
\end{eqnarray}
with dilogarithms $\Li(y^p)=\sum_{n>0} y^{p n}/n^2$, $p=1,2$, giving
Clausen's integral if $z\in[0,1]$.
\newpage
\section{Limiting cases}
For $z\to0$, one may approximate the coefficients by truncation of their
Taylor series
\begin{equation}
C^J(z)=\sum_{n=0}^\infty\left(a^J_n+\frac{\as}{\pi}c^J_n\right)z^n\,;\quad
b^J_n\equiv c^J_n/a^J_n\,,
\label{bj}
\end{equation}
with one-loop moments given by~\cite{RRY}
\begin{equation}
a_n\rV=-\frac{2n+2}{15}\frac{(4)_n}{(\frac72)_n}\,,\quad
a_n\rA=-\frac13\frac{(3)_n}{(\frac52)_n}\,,\quad
a_n\rS=-\frac{3n+4}{12}\frac{(2)_n}{(\frac52)_n}\,,\quad
a_n\rP=-\frac{n-4}{12}\frac{(2)_n}{(\frac32)_n}\,,
\label{aj}
\end{equation}
where $(a)_n=\Gamma(a+n)/\Gamma(a)$. The two-loop corrections, $b^J_n$, are
easily obtained from Table~1, using the expansions~\cite{SCG}
\begin{equation}
G(z)=\sum_{n=0}^\infty g_n z^n\,,\,
(1-z)G^2(z)=\sum_{n=0}^\infty\frac{g_n z^n}{n+1}\,,\,
H(z)=\sum_{n=0}^\infty\left(\sum_{r=0}^n\frac1{2r+1}
-\sum_{r=1}^n\frac3{2r}\right)g_n z^n\,,
\label{tay}
\end{equation}
with $g_n=n!/(3/2)_n$. Results for $n\leq10$ are given in Table~2, which
may be extended up to $n=200$ in a few minutes of CPUtime. Since
$a\rP_4=0$, $b\rP_4$ is undefined. We find that $c\rP_4=-3266/4725$. Two
features of Table~2 are notable: the relation $b\rP_1=b\rA_0$ ensures the
absence of a singularity in\Eq{pia} as $q^2\to0$; the scheme-independent
value $b\rP_0=b\rS_0=11/4$ gives, via a Ward identity~\cite{DJB,SCG}, the
new two-loop term in the heavy-quark expansion
\begin{equation}
\vev{m\overline\psi\psi}=-\df14\Pi\np\rP(0)=-\frac{\vev{\as G^2}}{12\pi}
\left(1+\frac{11}{4}\frac{\as}{\pi}+O(\as^2)\right)+O(1/m^2)\,,
\label{hqe}
\end{equation}
whose one-loop term was used in~\cite{G+B}. In the \MS\ scheme, at $\mu=m$,
one obtains two-loop corrections $\overline{b}^J_n=b^J_n-\frac43(2n+n_J)$.
Comparing Table~2 with the corresponding \MS\ tables in~\cite{PL2}, we find
that the latter give $(\overline{b}^J_n-2)$, instead of $\overline{b}^J_n$,
for the reasons given in the introduction.
\par
Next we consider the limit $z\to-\infty$, which yields the asymptotic
corrections
\begin{equation}
\frac{C\rV_2}{C\rV_1}\to5\,,\quad
\frac{C\rA_2}{C\rA_1}\to\frac{34}{9}\,,\quad
\frac{C\rS_2}{C\rS_1}\to6+2\ln\left(\frac{m^2}{-q^2}\right),\quad
\frac{C\rP_2}{C\rP_1}\to4+2\ln\left(\frac{m^2}{-q^2}\right).
\label{asy}
\end{equation}
To relate these results to previous work, we use the methods of~\cite{G+B}.
The vector and axial light-quark results of~\cite{LG2}, combined
with our heavy-quark result\Eq{hqe}, give
\begin{eqnarray}
q^4\Pi\np\rV(q^2)&\to&
\frac{\vev{\as G^2}}{12\pi}\left(1+\frac76\frac{\as}{\pi}\right)
+2\vev{m\overline\psi\psi}\left(1+\frac13\frac{\as}{\pi}\right)=
-\frac{\vev{\as G^2}}{12\pi}\left(1+5\frac{\as}{\pi}\right)\!,
\label{hqv}\\
q^2\Pi\np\rA(q^2)&\to&
\frac{\vev{\as G^2}}{12\pi}\left(1+\frac76\frac{\as}{\pi}\right)
-2\vev{m\overline\psi\psi}\left(1+\frac73\frac{\as}{\pi}\right)=
\frac{\vev{\as G^2}}{4\pi}\left(1+\frac{34}9\frac{\as}{\pi}\right)\!.\Pm
\label{hqa}
\end{eqnarray}
The (pseudo)scalar results of~\cite{LG2} similarly agree
with\Eqq{hqe}{asy}. Note, in\Eq{hqv}, the change in sign between one-loop
light- and heavy-quark $G^2$ terms. (The authors of~\cite{YF} regret that a
one-loop sign error in their previous work obscured the need to
use~\cite{G+B}.)
\par
Finally, for the threshold behaviour in the vector channel, as $z\to1$, we
obtain
\begin{equation}
C_2\rV(z)=
-\frac{\frac{197}{2304}\pi^2}{(1-z)^3}
+\frac{\frac{65}{768}\pi}{(1-z)^{5/2}}
-\frac{\frac{413}{6912}\pi^2}{(1-z)^2}
+\frac{\frac{17}{72}\pi\ln(1-z)}{(1-z)^{3/2}}
+O\left(\frac{1}{(1-z)^{3/2}}\right),
\label{thr}
\end{equation}
whose first term agrees with~\cite{NR}. (An input used in~\cite{YF} is
ruled out by the second term.)
\newpage
\section{Approximations}
We now assess previous numerical methods~\cite{AI93,YF,BFT,CONF}, by
testing the accuracy to which they predict new features of the exact result
for $C_2\rV(z)$. As input for the methods we take the following `old' data:
the {\em leading\/} asymptotic behaviour, from~\cite{LG2}; the {\em
leading\/} term in the threshold expansion, from~\cite{NR}; the radiative
corrections $\{b\rV_n|\,n<7\}$ to the first 7 moments. (This is the input
used in~\cite{YF}, corrected for errors in~\cite{YF,PL1,PL2}.)
\par
First we consider the approximate spectral ansatz of~\cite{YF}, which is
identical to using the following (exact) one-loop and (approximate)
two-loop threshold expansions:
\begin{equation}
C_1\rV(z)=-\df16F_2(z)+\df1{30}F_6(z)\,,\quad
C_2\rV(z)\approx\sum_{k=1}^{N+3}f_k F_k(z)\,;\quad
F_k(z)={}_2F_1(3,1;\df{k+1}2;z)\,,
\label{anz}
\end{equation}
with hypergeometric basis functions, $F_k(z)$, giving polynomials in
$1/(1-z)$, for $k=1,3,5$, and terms involving $G(z)$ or $\ln(1-z)$,
otherwise. The $(N+3)$ two-loop coefficients, $f_k$, are determined by the
first $N$ moments and by 3 further constraints~\cite{YF}:
\begin{equation}
f_1=-\frac{197\pi^2}{2304}\,,\quad
\sum_{k=2}^{N+3}(k-1)f_k=0\,,\quad
\sum_{k=2}^{N+3}k(k-1)f_k=\frac{10}3\,,
\label{cond}
\end{equation}
giving the leading threshold singularity, derived from~\cite{NR}, and
$z^2C\rV_2(z)\to-5/12$, derived from~\cite{LG2}, as $z\to-\infty$. We
remark that ansatz\Eq{anz} of~\cite{YF} cannot reproduce the form of the
4th term in the true threshold expansion\Eq{thr}. We assess it by comparing
the values of $f_{2,3}$, required by the input, with those required by the
second and third terms in\Eq{thr}, namely $f_2=65/144$ and
$f_3=-413\pi^2/3456$, and also by comparing the output for
$\{b\rV_n|\,n=N\ldots10\}$ and $B(z)=C\rV_2(z)/C\rV_1(z)$ with exact
results. Table~3 shows the very high accuracy achieved with 7 moments, in
all but the $f_{2,3}$ tests. The results using only 4 moments are almost as
good, except for $z\sim3$. Remarkably, the $N=1$ column shows that just 3
input numbers, namely $b\rV_0$, $f_1$ and $B(-\infty)$, give 10 additional
moments to better than 1\% accuracy, which is most encouraging for
three-loop applications.
\par
For the last 3 columns of Table~3, we mapped~\cite{CONF} and
Pad\'{e}-approximated~\cite{AI93,BFT,CONF}
\begin{equation}
z(1-z)^2C\rV_2(z)+\frac{5z}{12}-\frac{f_1}{1-z}
=\frac{D(\omega)}{1-\omega}\,;\quad\quad
z=\frac{4\omega}{(1+\omega)^2}\,,
\label{map}
\end{equation}
with the $z$\/-plane, cut along $z\in[1,\infty]$, mapped to the unit disk,
$|\omega|<1$. By construction, $D(\omega)$ is finite at $\omega=1$
(i.e.~$z=1$) and diverges only logarithmically as $\omega\to-1$
(i.e.~$z\to-\infty$). Its value at $\omega=0$ (i.e.~$z=0$) is $-f_1$. With
$N=1,4,7$ moments as input, we computed the $[0/1]$, $[2/2]$, $[3/4]$
Pad\'{e} approximants to $D(\omega)$, which were expanded to give the
further moments of Table 3. {From} the approximations to $D(1)$ and
$D^\prime(1)$, we obtained the second and third terms in the
Pad\'{e}-approximated threshold expansion and compared them with the exact
ones, in\Eq{thr}. Finally, we computed Pad\'{e}-approximated values of
$B(z)=C_2\rV(z)/C_1\rV(z)$, using~\cite{CONF} $\omega=\exp\left(2\ri\arccos
z^{-1/2}\right)$ in the cases with $z\in[1,\infty]$.
\par
Table~3 clearly demonstrates that both\Eq{anz} and\Eq{map} are highly
effective predictors of additional moments, even with {\em very\/} limited
input. More input is needed to achieve high accuracy on the cut, since
neither method performs well in the $f_{2,3}$ tests. With $N=10$ moments,
the Pad\'{e} method gives the modulus of $C_2\rV(z)$ to $0.05\%$ and the
phase to $0.04^\circ$, on the entire cut, whilst the figures for
ansatz\Eq{anz} are $0.5\%$ and $0.2^\circ$, respectively.
\newpage
\section{Summary and conclusions}
Our main analytical results are in~(\ref{vres}--\ref{hz}) and Table~1,
which give the two-loop coefficient functions of the gluon condensate
in the correlators of heavy-quark currents. As suggested in~\cite{AI92,BFT},
computer algebra in $d$ dimensions proves to be a most efficient way of
obtaining new 4-dimensional results, whose analytical complexity enters only
at the final stage, in this case via the dilogarithms of\Eq{hz}.
\par
By combining and refining the {\sc Reduce} programs developed
for~\cite{BFT,PL1}, we obtained a complete analytical result in the vector
channel in about 2 hours of CPUtime on a 486DX2-50 PC, whose timing in the
standard {\sc Reduce} test is 3~s. In comparison, it took 20 hours to obtain
merely the first 7 moments in~\cite{PL1}, using a machine with a benchmark
of 11~s. Our improvement in generality and speed entailed
considerable programming effort to produce efficient code, of the type
used in~\cite{BFT}, in order to process the many scalar integrals in the
0.7 Mbyte of output from the programs used in~\cite{PL1}.
This involved systematic implementation of recurrence relations for
$_5F_4$ hypergeometric functions, which was exhaustively checked
by independent {\sc Form} code, written for~\cite{BFT}.
\par
We have found full agreement with previous results in three limiting cases:
$z\to0$, $z\to-\infty$, and $z\to1$, after correcting errors
in~\cite{YF,PL1,PL2}. Using a Ward identity, we have obtained the new
two-loop term in the heavy-quark expansion\Eq{hqe}.
\par
Combining our new results with the procedure used in~\cite{YF},
to estimate the gluon condensate from empirical data~\cite{MOM},
we arrive at $\vev{(\as/\pi)G_{\mu\nu}^a G^{\mu\nu}_a}\approx0.021$~GeV$^4$,
as compared with 0.025~GeV$^4$ in~\cite{YF}. Our two-loop extraction
thus gives a value that is still about twice as large as the effective
one-loop value~\cite{LG2}.
\par
We have critically assessed two methods of numerical estimation:
the approximate spectral~\cite{YF} ansatz\Eq{anz}; and Pad\'{e}
approximation~\cite{BFT,CONF} after the mapping\Eq{map}.
Table~3 shows that each performs rather well, even with limited input.
Despite its analytical simplicity (rational approximation to $D(\omega)$
in\Eq{map}) the Pad\'{e} method is as good as\Eq{anz}
for moment prediction and better for achieving high accuracy on the cut.
We recommend both methods for future applications, with agreement between
them providing a useful cross check.
Note that the Pad\'{e} method is more general, as it does not rely
on the specific form of a lower-loop result. Most
importantly, one should use information from all 3 limiting cases,
in each method. We find that 4-figure accuracy can be obtained on the cut
using only 10 moments in the Pad\'{e} method of\Eq{map}, whilst
17 moments are needed to achieve comparable accuracy
using only the methods of~\cite{CONF}, without the limiting values as
$z\to1$ and $z\to-\infty$.
\par
In conclusion, we believe that our analytical progress and accuracy
of numerical approximation offer
real prospects of extending the two-loop approaches
of~\cite{BFT,DST,CONF} to {\em three\/} loops.
Future work can be based on calculation of a few three-loop terms as
$z\to-\infty$~\cite{GM2} and $z\to0$~\cite{3LP,GM2}, for both QED and QCD
two-point functions. In both theories, the leading three-loop term at
threshold is already known~\cite{NR}. We anticipate being able to obtain
an accurate three-loop approximation for the photon propagator, for use in
multi-loop QED~\cite{GM2}, and accurate estimates of three-loop corrections
to QCD sum rules. Finally we note that mapped Pad\'{e} methods
are well suited to a very wide variety of standard-model applications,
as indicated by their successful application to two-loop three-point
functions~\cite{AI93,CONF}.
\newpage\noindent{\em Acknowledgments} We thank Denis Perret-Gallix for his
part in organizing AI-HENP~92~\cite{AI92}, which resulted in our
international ASTEC collaboration, Karl-Heinz Becks for his part in
organizing AI-HENP~93~\cite{AI93}, which assisted in the planning of this
work, and Tony Hearn, for adapting {\sc Reduce}~3.5~\cite{RED}, to suit the
needs of our programs.
\vfill
\begin{center}Table~1: The polynomial coefficients of\Eq{pj},
for use in~(\ref{vres}-\ref{pres})\end{center}
$\qP\rV_1(z)=-\frac1{72}(613+8828z-22092z^2+13248z^3)$\\[3pt]
$\qP\rV_2(z)=\pfrac1{72}(659-3774z+30184z^2-55344z^3+26496z^4)$\\[3pt]
$\qP\rV_3(z)=-\frac1{36}(23+1456z-41064z^2+195024z^3-350688z^4
             +275328z^5-79488z^6)$\\[3pt]
$\qP\rV_4(z)=\pfrac49z(473-3614z+6504z^2-3312z^3)$\\[7pt]
$\qP\rA_1(z)=\pfrac1{216}(1241-8390z+6624z^2)$\\[3pt]
$\qP\rA_2(z)=-\frac1{216}(1339+4128z-17656z^2+13248z^3)$\\[3pt]
$\qP\rA_3(z)=\pfrac1{108}(49-2404z+32328z^2-98112z^3+107616z^4
             -39744z^5)$\\[3pt]
$\qP\rA_4(z)=\pfrac4{27}z(467-2000z+1656z^2)$\\[7pt]
$\qP\rS_1(z)=-\frac1{24}(77+2402z-2208z^2)$\\[3pt]
$\qP\rS_2(z)=\pfrac1{24}(187-1636z+5416z^2-4416z^3)$\\[3pt]
$\qP\rS_3(z)=-\frac1{12}(55+82z-8880z^2+30624z^3-35040z^4
             +13248z^5)$\\[3pt]
$\qP\rS_4(z)=\pfrac43z(121-632z+552z^2)$\\[7pt]
$\qP\rP_1(z)=-\frac1{24}(3213-9542z+6624z^2)$\\[3pt]
$\qP\rP_2(z)=\pfrac1{24}(2739+9428z-26104z^2+13248z^3)$\\[3pt]
$\qP\rP_3(z)=-\frac1{12}(99-11470z+73008z^2-151776z^3+130080z^4
             -39744z^5)$\\[3pt]
$\qP\rP_4(z)=\pfrac43(54-1299z+2936z^2-1656z^3)$
\vfill
\begin{center}Table~2: Values of $b^J_n=c^J_n/a^J_n$ for
$J=\VASP$ and $n\leq10$\end{center}
\[\begin{array}{cllll}
\quad n\quad&{\rm V}&{\rm A}&{\rm S}&{\rm P}\\[5pt]
0&
{1469\over162}&
{131\over24}&
{11\over4}&
{11\over4}\\[3pt]
1&
{135779\over12960}&
{3193\over486}&
{2447\over504}&
{131\over24}\\[3pt]
2&
{1969\over168}&
{72631\over9720}&
{2813\over450}&
{307\over45}\\[3pt]
3&
{546421\over42525}&
{124847\over15120}&
{160591\over21840}&
{97073\over15120}\\[3pt]
4&
{661687433\over47628000}&
{36591619\over4082400}&
{400895\over48384}&
\mbox{undefined}\\[3pt]
5&
{1800293669\over121080960}&
{604286927\over62868960}&
{1457\over160}&
{28678327\over1663200}\\[3pt]
6&
{180657583657\over11442150720}&
{5566166237\over544864320}&
{16727257\over1698840}&
{1854828721\over113513400}\\[3pt]
7&
{72346752463\over4341887550}&
{26439650563\over2451889440}&
{398248999\over37837800}&
{72267337\over4324320}\\[3pt]
8&
{377046256819\over21549939840}&
{9437661186719\over833642409600}&
{321477183403\over28817268480}&
{5495760533\over316673280}\\[3pt]
9&
{22322605695461\over1220106888000}&
{98175636210793\over8296726838400}&
{30518020922371\over2597964969600}&
{159068122366537\over8799558768000}\\[3pt]
10&
{162570424650982\over8527479553593}&
{38340293613343\over3111272564400}&
{2191394981453\over178086308400}&
{28163524523\over1496523600}
\end{array}\]
\newpage
\begin{center}Table~3: Assessment of ansatz\Eq{anz} and mapping\Eq{map},
using $N$ moments\end{center}
\[\begin{array}{|c|c|ccc|ccc|}\hline
\mbox{test}&\mbox{exact}&
\mbox{ansatz}&\mbox{ansatz}&\mbox{ansatz}&
\mbox{mapped}&\mbox{mapped}&\mbox{mapped}\\
\mbox{feature}&\mbox{result}&
N=7&N=4&N=1&
N=7&N=4&N=1\hl
b\rV_{0}  &9.06790&\noshow&\noshow&\noshow&\noshow&\noshow&\noshow\\
b\rV_{1}  &10.4768&\noshow&\noshow&10.4484&\noshow&\noshow&10.4778\\
b\rV_{2}  &11.7202&\noshow&\noshow&11.6675&\noshow&\noshow&11.7281\\
b\rV_{3}  &12.8494&\noshow&\noshow&12.7761&\noshow&\noshow&12.8666\\
b\rV_{4}  &13.8928&\noshow&13.8928&13.8019&\noshow&13.8927&13.9203\\
b\rV_{5}  &14.8685&\noshow&14.8684&14.7625&\noshow&14.8682&14.9065\\
b\rV_{6}  &15.7888&\noshow&15.7885&15.6695&\noshow&15.7881&15.8369\\
b\rV_{7}  &16.6625&16.6625&16.6620&16.5315&16.6625&16.6614&16.7204\\
b\rV_{8}  &17.4964&17.4964&17.4955&17.3549&17.4964&17.4946&17.5636\\
b\rV_{9}  &18.2956&18.2956&18.2944&18.1446&18.2956&18.2931&18.3715\\
b\rV_{10} &19.0643&19.0643&19.0626&18.9046&19.0643&19.0609&19.1485\hl
   -f_1   &0.84388&\noshow&\noshow&\noshow&\noshow&\noshow&\noshow\\
\Pm f_2   &0.45139&0.45868&0.50439&0.57116&0.47972&0.53134&0.40795\\
   -f_3   &1.17944&1.49727&2.09688&2.23783&1.86522&2.39296&1.62672\hl
B(-\infty)&5.00000&\noshow&\noshow&\noshow&\noshow&\noshow&\noshow\\
B(-9)     &5.58641&5.58639&5.58495&5.62888&5.58733&5.58208&5.61983\\
B(-3)     &6.28843&6.28843&6.28800&6.33614&6.28847&6.28730&6.31060\\
B(-0.9)   &7.41565&7.41565&7.41561&7.44826&7.41565&7.41557&7.42198\\
B(-0.3)   &8.29735&8.29735&8.29735&8.31290&8.29735&8.29734&8.29824\\
B(0.3)    &10.3704&10.3704&10.3704&10.3449&10.3704&10.3704&10.3745\\
B(0.9)    &25.6939&25.6939&25.6835&25.4851&25.6928&25.6745&25.8290\\
|B(3)|    &1.81071&1.80980&1.91045&2.47868&1.80718&2.00465&2.21432\\
|B(9)|    &4.47078&4.47366&4.44784&4.46713&4.46202&4.43571&4.69021\\
\hline\end{array}\]
\newpage
\raggedright


\begin{thebibliography}{99}
\bibitem{AI92}
D.J.\ Broadhurst,
in {\sl New computing techniques in physics research II\/},\\
ed.\ D.\ Perret-Gallix (World Scientific, Singapore, 1992) p.\ 579.
\bibitem{AI93}
J.\ Fleischer,
in {\sl New computing techniques in physics research III\/},\\
ed.\ K.-H.\ Becks and D.\ Perret-Gallix
(World Scientific, Singapore, 1994) p.\ 551;\\
D.J.\ Broadhurst, {\sl ibid} p.\ 511.
\bibitem{YF}
P.A.\ Baikov, V.A.\ Ilyin and V.A.\ Smirnov,
{\sl Phys.At.Nucl.}\ {\bf B56}(11) (1993) 1527.
\bibitem{BFT}
D.J.\ Broadhurst, J.\ Fleischer and O.V.\ Tarasov,
{\sl Z.Phys.}\ {\bf C60} (1993) 287.
\bibitem{PL1}
K.G.\ Chetyrkin, V.A.\ Ilyin, V.A.\ Smirnov and A.Yu.\ Taranov,\\
{\sl Phys.Lett.}\ {\bf B225} (1989) 411.
\bibitem{PL2}
P.A.\ Baikov, K.G.\ Chetyrkin, V.A.\ Ilyin, V.A.\ Smirnov
and A.Yu.\ Taranov,\\
{\sl Phys.Lett.}\ {\bf B263} (1991) 481.
\bibitem{MAS}
D.J.\ Broadhurst,
{\sl Z.Phys.}\ {\bf C47} (1990) 115.
\bibitem{B+R}
R.\ Barbieri and E.\ Remiddi,
{\sl Nuovo Cimento} {\bf 13} (1973) 99.
\bibitem{LG2}
M.A.\ Shifman, A.I.\ Vainshtein and V.I.\ Zakharov,
{\sl Nucl.Phys.}\ {\bf B147} (1979) 385;\\
P.\ Pascual and E.\ de Rafael,
{\sl Z.Phys.}\ {\bf C12} (1982) 127;\\
K.G.\ Chetyrkin, S.G.\ Gorishny and V.P.\ Spiridonov,
{\sl Phys.Lett.}\ {\bf B160} (1985) 149;\\
G.T.\ Loladze, L.R.\ Surguladze and F.V.\ Tkachov,
{\sl Phys.Lett.}\ {\bf B162} (1985) 363;\\
L.R.\ Surguladze and F.V.\ Tkachov,
{\sl Nucl.Phys.}\ {\bf B331} (1990) 35.
\bibitem{NR}
M.B.\ Voloshin,
{\sl Sov.J.Nucl.Phys.}\ {\bf 36} (1982) 143.
\bibitem{DST}
A.I.\ Davydychev and J.B.\ Tausk,
{\sl Nucl.Phys.}\ {\bf B397} (1993) 123;\\
A.I.\ Davydychev, V.A.\ Smirnov and J.B.\ Tausk,
{\sl Nucl.Phys.}\ {\bf B410} (1993) 325.
\bibitem{CONF}
J.\ Fleischer and O.V.\ Tarasov,
BI--TP/93--78 (1993), hep-ph/9403230.
\bibitem{3LP}
D.J.\ Broadhurst,
{\sl Z.Phys.}\ {\bf C54} (1992) 599.
\bibitem{GM2}
D.J.\ Broadhurst, A.L.\ Kataev and O.V.\ Tarasov,
{\sl Phys.Lett.}\ {\bf B298} (1993) 445;\\
T.\ Kinoshita,
{\sl Phys.Rev.}\ {\bf D47} (1993) 5013.
\bibitem{DJB}
D.J.\ Broadhurst,
{\sl Phys.Lett.}\ {\bf B101} (1981) 423.
\bibitem{SCG}
S.C.\ Generalis,
Open University thesis, OUT--4102--13 (1984).
\bibitem{RED}
A.C.\ Hearn,
{\sc Reduce} user's manual, version 3.5, Rand publication CP78 (1993).
\bibitem{RRY}
L.J.\ Reinders, H.R.\ Rubinstein and S.\ Yazaki,
{\sl Phys.Lett.}\ {\bf B94} (1980) 203;
{\sl Phys.Rep.}\ {\bf 127} (1985) 1.
\bibitem{G+B}
S.C.\ Generalis and D.J.\ Broadhurst,
{\sl Phys.Lett.}\ {\bf B139} (1984) 85;\\
D.J.\ Broadhurst and S.C.\ Generalis,
{\sl Phys.Lett.}\ {\bf B142} (1984) 75.
\bibitem{MOM}
G.\ Grunberg,
{\sl Acta.Phys.Pol.}\ {\bf B16} (1985) 491.
\end{thebibliography}
\end{document}